\documentclass[letterpaper,twocolumn,superscriptaddress,
unpublished,allowtoday]{quantumarticle}
\pdfoutput=1
\usepackage[utf8]{inputenc}
\usepackage[english]{babel}
\usepackage[T1]{fontenc}
\usepackage{amsmath}
\usepackage{hyperref}

\usepackage{tikz}
\usepackage{lipsum}

\usepackage[numbers,sort&compress]{natbib}

\newtheorem{definition}{Definition}

\newcommand*\sectionname{Section}
\newcommand*\equationname{Equation}

\newcommand{\sectionref}[1]{\sectionname~\ref{#1}}
\newcommand{\figureref}[1]{\figurename~\ref{#1}}
\newcommand{\tableref}[1]{\tablename~\ref{#1}}
\newcommand{\equationref}[1]{\equationname~\ref{#1}}

\begin{document}

\title{Stochastic search for approximate compilation of unitaries}
\date{January 13, 2021}
\author{Ryan Shaffer}
\email{ryan.shaffer@berkeley.edu}
\orcid{0000-0002-3196-4529}
\affiliation{Department of Physics, University of California, Berkeley, CA, USA}

\maketitle

{\color{red} NOTE: This manuscript has been superseded by \href{https://arxiv.org/abs/2205.13074}{\color{red} https://arxiv.org/abs/2205.13074}.}
\vspace{2mm}

\begin{abstract}
    Compilation of unitaries into a sequence of physical quantum gates is a critical prerequisite for execution of quantum algorithms.
    This work introduces STOQ, a stochastic search protocol for approximate unitary compilation into a sequence of gates from an arbitrary gate alphabet.
    We demonstrate STOQ by comparing its performance to existing product-formula compilation techniques for time-evolution unitaries on system sizes up to eight qubits.
    The compilations generated by STOQ are less accurate than those from product-formula techniques, but they are similar in runtime and traverse significantly different paths in state space.
    We also use STOQ to generate compilations of randomly-generated unitaries, and we observe its ability to generate approximately-equivalent compilations of unitaries corresponding to shallow random circuits.
    Finally, we discuss the applicability of STOQ to tasks such as characterization of near-term quantum devices.
\end{abstract}

\section{Introduction}

A critical prerequisite to executing any algorithm on a physical quantum computer is the process commonly known as quantum compilation.
   One of the primary tasks of quantum compilation is the conversion of a target unitary operation into a sequence of quantum gates that are native to the physical device being used \cite{Barenco1995ElementaryComputation, Cybenko2001ReducingOperations, Harrow2002EfficientGates, Javadiabhari2015ScaffCC:Programs}.
   Because unitary operators belong to a continuous space, such compilation in general results in gate sequences which are only approximately equivalent to the target unitary.
   For example, one of the earliest quantum compilation techniques, the Solovay-Kitaev method \cite{Dawson2006TheAlgorithm}, compiles gate sequences that differ from the target unitary by an amount that can be made as small as desired.

Traditional compilation, both in the classical and quantum realms, is most often a deterministic process, using rules and heuristics to efficiently synthesize a desired program from
   the native assembly instructions (in classical compilation)
   or native physical gates (in quantum compilation).
But in some cases, adding stochasticity to the compilation process has been shown to produce advantages in the resulting program.
   In classical compilation, a technique known as stochastic superoptimization \cite{Schkufza2013StochasticSuperoptimization} has been shown in certain cases to produce significantly shorter programs than the best-in-class compilers and optimizers.
   In quantum compilation, techniques such as randomized compiling \cite{Wallman2016NoiseCompiling} have been demonstrated to improve  noise resilience by depolarizing errors that occur during program execution.

In the field of quantum compilation, special attention has been paid to compilation of unitaries which result from the time-evolution of physically-realizable Hamiltonians.
The compiled sequences in these cases can be executed to perform what is known as ``Hamiltonian simulation'', or more broadly, ``quantum simulation''.
Such approaches are of special interest in fields such as quantum chemistry, where it is desirable to use a quantum computer to simulate the dynamics of physical systems.
Common approaches to this problem include
   product formula techniques such as the Suzuki-Trotter decomposition \cite{Hatano2005FindingOrders}
   and qubitization \cite{Low2019HamiltonianQubitization},
which deterministically compile the time-evolution unitary for a given Hamiltonian into a sequence of quantum gates.

Approaches involving stochasticity have recently been shown to be advantageous in some cases.
   Adding randomization to the
    Suzuki-Trotter decomposition \cite{Childs2019FasterRandomization}
    creates approximate compilations that are better both theoretically and empirically.
   A stochastic compilation protocol known as
    QDRIFT \cite{Campbell2019RandomSimulation},
    where gate probabilities are weighted according to the strength of each term in the Hamiltonian rather than using a product formula directly,
    has been shown to produce much more efficient compilations in many cases.
   An interpolation of these two methods called
    SparSto \cite{Ouyang2020CompilationSparsification}
    has also been proposed, which takes some of the advantages of each method.
The efficiency of these compilation methods is generally independent of system size when applied to problems involving sparse Hamiltonians.

However, efficiency is not the only attribute of a quantum compilation protocol that may be desired.
   In particular, increasing the randomness with which the quantum program is generated may be beneficial for purposes such as characterizing and benchmarking the resilience of a physical system to various types of noise.

This work introduces a stochastic approximate quantum unitary compilation scheme, abbreviated as STOQ, which uses a randomized search process to generate gate sequences that approximately implement a target unitary in terms of any arbitrary set of native gates.
   First, the technical implementation details of STOQ are described, along with some potential advantages and disadvantages.
   Next, we report results of applying STOQ to Hamiltonian time-evolution unitaries, where we compare its performance to existing methods on various metrics.
   We also demonstrate the use of STOQ to approximately compile gate sequences for randomly-generated unitaries.
   Finally, we discuss potential applications of STOQ, particularly for characterization and benchmarking tasks, and discuss avenues for future work and improvements to the technique.

\section{STOQ: A stochastic search protocol for approximate unitary compilation}\label{sec:stoq}

\subsection{Definitions}

The process of compilation requires specification of
    the unitary operation to be compiled,
    as well as the set of gates which are allowed to be used in the final compiled sequence.

\begin{definition}[Target unitary]
    For an $n$-qubit system, the target unitary is the
        $2^n$-dimensional unitary operator $U$
        implementing some desired effect on the system.
\end{definition}

The set of gates used for the compilation may, in general, be fixed or parameterized.

\begin{definition}[Fixed gates]
    Fixed gates, such as Cliffords, are discrete operations that can be represented by a fixed unitary matrix.
\end{definition}

\begin{definition}[Parameterized gates]
    Parameterized gates, such as rotations, are continuous operations that can be represented by a unitary matrix with one or more continuously-variable parameters.
\end{definition}

The allowed set of gates for the compilation may then consist of some combination of
    fixed and parameterized gates.

\begin{definition}[Gate alphabet]
    For an $n$-qubit system, the gate alphabet is a set of fixed gates and/or
    parameterized gates that represent the fundamental set of operations that
    can be physically applied to the system.
    Also called ``native gate set''.
\end{definition}

The problem of approximate quantum unitary compilation can now be stated as follows:

\begin{definition}[Approximate compilation]
    Given a target unitary $U$ and a gate alphabet $G$,
       find a sequence of gates $\{G_1, \dots, G_M\}$
          such that the product $G_M G_{M-1} \cdots G_1$ is approximately equivalent to $U$.
\end{definition} 

That is, given some appropriate distance metric
  which defines a distance $d$ between
    the sequence product $G_M G_{M-1} \cdots G_1$
    and the target unitary $U$,
  the compilation process treats $d$ as the value of a cost function to be minimized.

\subsection{Protocol description}\label{sec:protocol-description}

We now introduce STOQ, a stochastic protocol for solving the problem of approximate quantum unitary compilation.
We note that this work builds on and generalizes a similar technique used
   for variational quantum compilation algorithms
      \cite{Khatri2019Quantum-assistedCompiling, Sharma2020NoiseCompiling}.
At a high level, the protocol proceeds according to the pseudocode
displayed in \figureref{fig:stoq-algorithm}.

\begin{figure}
\begin{verbatim}
function StochasticCompilation
(params U, G, num_iterations):
  sequence := []
  beta := 0
  cost := Cost(U, Prod(sequence))
  for i in 1 to num_iterations:
    beta := IncreaseBeta(beta)
    new_sequence := RandomChange(sequence, G)
    new_cost := Cost(U, Prod(new_sequence))
    if Accept(cost, new_cost, beta):
      sequence := new_sequence
      cost := new_cost
  return sequence
\end{verbatim}
    \caption{
        Pseudocode for STOQ algorithm. 
        The inputs to the algorithm are
            the target unitary \texttt{U},
            the parameterized gate alphabet \texttt{G},
            and the number of iterations to perform \texttt{num\_iterations}.
        The algorithm is described in \sectionref{sec:stoq}.
    }
    \label{fig:stoq-algorithm}
\end{figure}

Intuitively, the STOQ algorithm can be thought of as
   a randomized exploration of the full space of possible $n$-qubit unitary operators
      (or the subspace that can be generated by $G$, if $G$ is not a universal gate set),
      using a technique known as Markov chain Monte Carlo (MCMC) search \cite{Hastings1970MonteApplications}.
   The algorithm is always initialized with an empty sequence, meaning that it always starts from the identity operator in the search space.
   At each iteration, a random step is proposed, in which an item is either added to or removed from the sequence.
      If this step brings the product of the sequence closer to the target unitary as determined by the cost function, it is accepted;
         otherwise, it is either accepted or rejected with some probability,
         where the probability of accepting such ``bad'' steps decreases with each iteration.
    The algorithm continues until some maximum number of iterations is reached,
        or alternatively, until the cost has reached a desired threshold.

One critical component of the algorithm is the choice of an appropriate and efficient cost function. Naturally, the cost function should be a distance measure between the the target unitary $U$ and the unitary $V$ which is the product of the currently-compiled sequence.
   One commonly-used and operationally-relevant choice is the \textit{trace distance}
      \begin{equation}
        D_{\textrm{trace}}(U,V)
        = \dfrac{1}{2} \textrm{Tr} \left\vert U-V \right\vert 
        ,
      \end{equation}
   but this is computationally expensive to compute for even moderately-sized $n$ due to the required diagonalization.
   A more efficient alternative, used also in variational quantum compilation approaches \cite{Khatri2019Quantum-assistedCompiling, Sharma2020NoiseCompiling},
   is the \textit{Hilbert-Schmidt distance}
      \begin{equation}\label{eq:hilbert-schmidt-distance}
        D_{\textrm{HS}}(U,V)
        = \left\vert \textrm{Tr} ( V^\dag U ) \right\vert
        ,
      \end{equation}
   which is operationally related to the fidelity of a process \cite{Nielsen2002AOperation}
   and can be shown in certain cases to be closely related to the trace distance \cite{Coles2019StrongStates}.
   For computational efficiency, then, we use a Hilbert-Schmidt cost function
      \begin{equation}\label{eq:cost}
        \texttt{Cost}(U,V)
        = 1 - \frac{1}{2^{n}} D_{\textrm{HS}}(U,V)
        ,
      \end{equation}
   noting that $\texttt{Cost}(U,V)$ ranges from 0 to 1 and vanishes if and only if $U$ and $V$ are equivalent up to a global phase.

\subsection{Implementation notes}\label{sec:protocol-implementation}

This section fills in a few important details of the STOQ protocol implementation,
    referring to the pseudocode representation in 
    \figureref{fig:stoq-algorithm}.
    
The compiled sequence is stored in the \texttt{sequence} variable,
    which is initially empty.
    The \texttt{RandomChange} function returns a modified sequence on each iteration, either
        by adding a randomly-drawn gate to the sequence
            from the parameterized gate alphabet \texttt{G}
            with randomly-generated parameter values,
        or by removing a gate from the sequence.
    The \texttt{Prod} function calculates the unitary that represents the product of all of the operations in the sequence,
    and the \texttt{Cost} function is implemented as described in \equationref{eq:cost}.

The variable \texttt{beta} is used as an annealing parameter for the compilation process.
    The function \texttt{IncreaseBeta} returns a slightly increased value of \texttt{beta} on each iteration.
    Defining the annealing parameter as
            $\beta = \texttt{beta}$
        and the cost difference of such a proposed change as
            $\Delta = \texttt{new\_cost} - \texttt{cost}$,
    the \texttt{Accept} function calculates the probability of accepting a proposed change as
    \begin{equation}
        P_\textrm{accept} =
            \begin{cases}
                e^{-\beta \Delta}   &\Delta > 0     \\
                1                   &\Delta \le 0
                .
            \end{cases}
    \end{equation}
    The probability of accepting ``bad'' proposed changes where the cost increases (i.e., where $\Delta > 0$) approaches zero as $\beta$ increases.

\section{Results}

\subsection{Compilation of time-evolution unitaries}\label{sec:compilation-time-evolution}

\begin{figure*}[ht!]
    \centering
    \includegraphics[width=\linewidth]{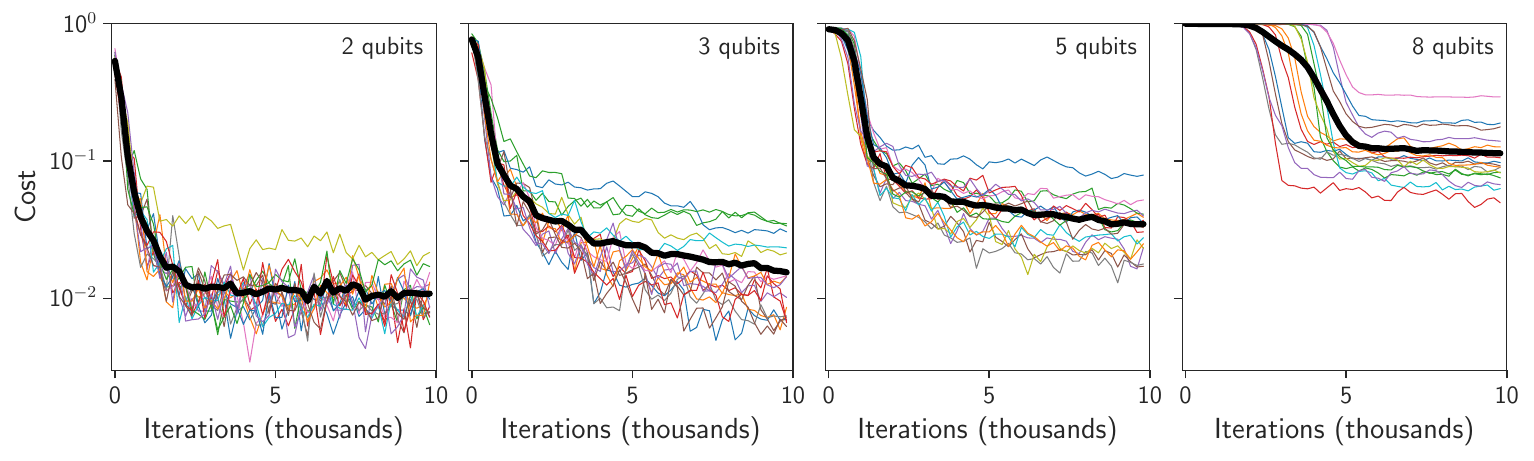}
    \caption{
        Compilation via STOQ for two-qubit, three-qubit, five-qubit, and eight-qubit versions of the time-evolution unitary
            from \equationref{eq:time-evolution-unitary}.
        Each of the 16 thin curves shows the value of the cost function from \equationref{eq:cost}
            during a single compilation using 10,000 iterations.
        The thick curve is the average of all runs.
    }
    \label{fig:time-evolution-unitary-compilation}
\end{figure*}

To demonstrate a simple application of STOQ,
    we choose an Ising-type Hamiltonian
        with nearest-neighbor coupling and transverse field
    \begin{equation}\label{eq:hamiltonian}
        H = \sum_{<i,j>} J_{ij} \sigma_x^{(i)}\sigma_x^{(j)}
            + \sum_{i} h_i \sigma_y^{(i)}
    \end{equation}
    where the coefficients $J_{ij}$ and $h_i$ are energies with arbitrarily-chosen values for each system size, as shown in \tableref{tab:coefficient-values}.

\begin{table}
    \centering
    \setlength\tabcolsep{5pt}
    \begin{tabular}{c | c | c | c | c | c | c | c}
        $n$ & $J_{12}$ & $J_{23}$ & $J_{34}$ & $J_{45}$ & $J_{56}$ & $J_{67}$ & $J_{78}$ \\
        \hline
        2   & 1.27     &          &          &          &          &          &        \\
        3   & 1.81     & 1.27     &          &          &          &          &        \\
        5   & 1.20     & 1.40     & 1.60     & 1.80     &          &          &        \\
        8   & 1.20     & 1.30     & 1.40     & 1.50     & 1.60     & 1.70     & 1.80   \\
    \end{tabular}
        
    \begin{tabular}{c | c | c | c | c | c | c | c | c}
        \hline\hline
        $n$ & $h_1$ & $h_2$ & $h_3$ & $h_4$ & $h_5$ & $h_6$ & $h_7$ & $h_8$ \\
        \hline
        2   & 1.54  & 1.19  &       &       &       &       &       &       \\
        3   & 1.54  & 1.19  & 0.53  &       &       &       &       &       \\
        5   & 1.60  & 1.30  & 1.00  & 0.70  & 0.40  &       &       &       \\
        8   & 1.40  & 1.10  & 0.80  & 1.00  & 1.20  & 1.50  & 1.70  & 1.30  \\
    \end{tabular}
    \caption{\label{tab:coefficient-values}
        Coefficients used for application of STOQ to the
            $n$-qubit Ising model Hamiltonian in \equationref{eq:hamiltonian}.
            Values are energies in arbitrary units where $\hbar = 1$.
    }
\end{table}

We then define the time-evolution unitary as $U_t(\tau) = e^{i H \tau}$, where we choose units such that $\hbar = 1$, and we concretely choose $\tau = 0.5$, such that
    \begin{equation}\label{eq:time-evolution-unitary}
        U = U_t(0.5) = e^{i H (0.5)}
    \end{equation}
    is the target unitary for compilation.

To apply STOQ, we need also to choose a parameterized gate alphabet $G$ from which to approximately compile a sequence.
In a physical device, it is often the case that the dynamics are implemented such that each term in $H$ can be individually controlled.
To define $G$ for such a device, we express the Hamiltonian as 
        $H = \sum_k H_k$,
    where each $H_k$ is one of the
        $\sigma_x\sigma_x$ or $\sigma_y$
        terms from \equationref{eq:hamiltonian},
    and choose
    \begin{equation}
        G = \bigcup_k \left\lbrace e^{i H_k t} \right\rbrace
        \quad -\epsilon \tau \le t \le \epsilon \tau
    \end{equation}
    where the allowed range for $t$ is chosen such that each gate is relatively short in comparison to the timescale of the dynamics of $H$. (In this demonstration we use $\epsilon = 0.2$.) 
    Negative times correspond to reversing the sign of the coefficient of a given term.
    
We then apply STOQ to compile many sequences that approximately implement $U$, using two-qubit, three-qubit, five-qubit, and eight-qubit versions of the corresponding Hamiltonian. 
    \figureref{fig:time-evolution-unitary-compilation} reports the cost for 16 such compilations as a function of the number of iterations.
    (Each run of 10,000 iterations for the five-qubit system takes around 15 minutes to complete on a typical desktop computer.)
    We observe that the stochastic search process rapidly reduces the cost at first before noticeably leveling off. For the two-qubit and three-qubit systems, this cost approaches a limit near $10^{-2}$ after 10,000 iterations. For the larger systems, the final average cost is higher, although even for the eight-qubit system, the final cost reaches a value below $10^{-1}$ for some compilations.

\begin{figure*}[ht!]
    \centering
    \includegraphics[width=\linewidth]{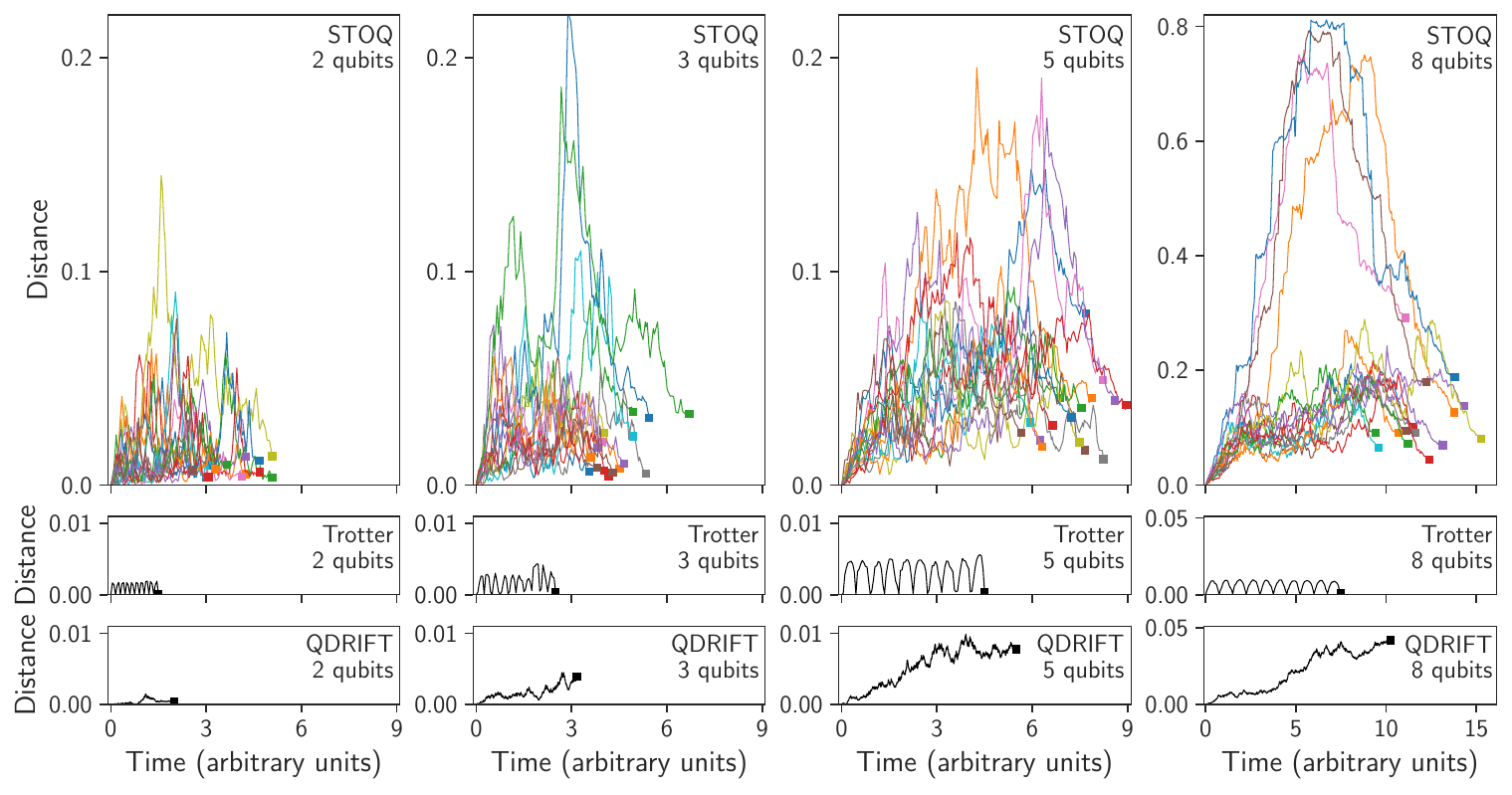}
    \caption{
        Distance from ideal path to compiled path,
            as defined in \equationref{eq:path-distance},
            for the time-evolution unitary from \equationref{eq:time-evolution-unitary}.
            Results are shown for 2-qubit, 3-qubit, 5-qubit, and 8-qubit
                implementations of the Ising model Hamiltonian from
                \equationref{eq:hamiltonian}.
            Each curve represents the execution of one compiled sequence.
            Filled squares are used to plot the
                overall running time of the compiled sequence and
                final cost of each compilation.
        Top row depicts the execution of 16 independent STOQ compilations,
            each using 10,000 iterations.
            Each curve corresponds to a curve of the same color
                in \figureref{fig:time-evolution-unitary-compilation}.
        Middle row depicts the execution of a typical randomized Suzuki-Trotter compilation using 10 steps.
        Bottom row depicts the execution of a typical QDRIFT compilation using 1,000 repetitions.
    }
    \label{fig:compilation-distance}
\end{figure*}

To compare STOQ to existing compilation techniques, we also compile sequences to approximately implement $U$ using
        the randomized Suzuki-Trotter decomposition \cite{Childs2019FasterRandomization}
        and the QDRIFT stochastic compilation protocol \cite{Campbell2019RandomSimulation}.
    STOQ is designed to create more randomness in the resulting path taken through state space.
        To compare these paths quantitatively, we choose to compare the various methods to an ideal version where $H$ is directly implemented for time $\tau$.
        We define the \textit{ideal path} as the path taken by this ideal time evolution, and we define the \textit{compiled path} as the path taken by the compiled sequence, which we represent as a sequence of gates $\{ G_1, \dots, G_M \}$.
        We then calculate the path distance $d_m$ from the ideal path to step $m$ of the compiled path, where $1 \le m \le M$, as
        \begin{equation}\label{eq:path-distance}
            d_m = \min_{t\,\in\,[0, \tau]}
                D_{\textrm{HS}} \left(
                    e^{i H t}, \ G_m G_{m-1} \cdots G_1 \right)
            ,
        \end{equation}
        where $D_{\textrm{HS}}$ is the Hilbert-Schmidt distance defined in \equationref{eq:hilbert-schmidt-distance}. Thus $d_m$ is the shortest distance from step $m$ of the compiled path to any point in the ideal path.

Results for each compilation technique are shown
    in \figureref{fig:compilation-distance},
    and statistics for the five-qubit example
    are displayed in \tableref{tab:compilation-distance}.
We observe that the STOQ compilations result in a significantly greater path distance from the ideal evolution than the other approaches,
    and that the total running time of the compiled sequence resulting from the various compilations is within a factor of two.

However, the final cost of the STOQ compilations is typically at least an order of magnitude larger than the compilations created using the randomized Suzuki-Trotter and QDRIFT techniques, both of which can reach arbitrarily low costs by increasing the number of steps. This implies that STOQ would not be a useful tool for applications that require high-fidelity compilations.

\begin{table}[b!]
    \centering
    \begin{tabular}{c | c | c | c | c}
                  & Ideal        & Trotter   & QDRIFT   & STOQ    \\ \hline
        Time      & 0.50         & 4.50      & 5.50     & 7.32    \\
        Mean($d$) & \textemdash  & 0.0032    & 0.0053   & 0.0469  \\
        Max($d$)  & \textemdash  & 0.0056    & 0.0099   & 0.1133  \\
        Cost      & \textemdash  & 0.0003    & 0.0077   & 0.0328
    \end{tabular}
    \caption{\label{tab:compilation-distance}
        Statistics resulting from various compilations of the five-qubit time-evolution unitary
            from \equationref{eq:time-evolution-unitary},
        where the ideal evolution occurs for $\tau=0.5$.
        Average total running time of the compiled sequence,
            average distance Mean($d$),
            maximum distance Max($d$),
            and final cost
        are listed for each of the compilation techniques.
        Corresponds to five-qubit plots in \figureref{fig:compilation-distance}.
    }
\end{table}

\subsection{Compilation of random unitaries}\label{sec:compilation-random-unitaries}

\begin{figure*}[ht]
    \centering
    \includegraphics[width=0.99\linewidth]{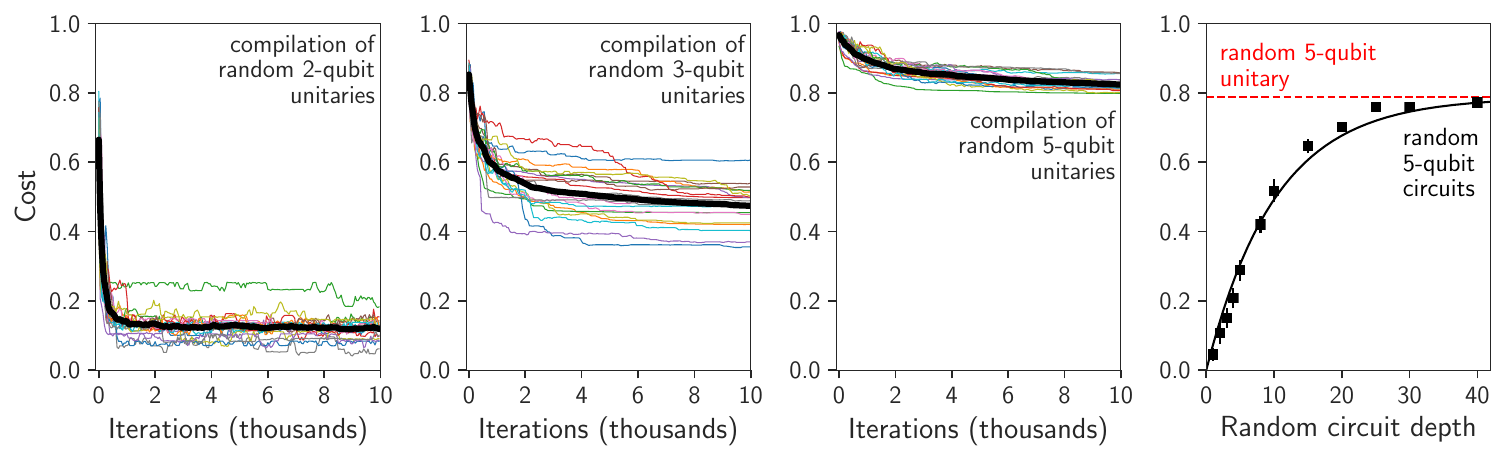}
    \caption{
        Compilation via STOQ of randomly-generated unitaries.
        The left three plots show the cost during the STOQ compilation process for randomly-generated 2-qubit, 3-qubit, and 5-qubit target unitaries.
        Each of the 20 thin curves shows the value of the cost function from \equationref{eq:cost}
            during a single compilation using 10,000 iterations.
            The thick curve is the average of all runs.
        The rightmost plot shows the final cost of the STOQ compilation for target unitaries generated by creating random 5-qubit circuits of varying average circuit depth.
            Circuit depth is calculated as the total number of gates divided by the number of qubits.
            Each point is the average of 20 compilations using 100,000 iterations.
            Error bars indicate standard error of the mean.
            The solid line is an exponential decay fit with one free parameter.
            The dashed line represents the average final cost of compiling a randomly-generated 5-qubit unitary.
    }
    \label{fig:random-unitary-compilation}
\end{figure*}

In addition to being used for sparse or highly structured unitaries such as those generated from Hamiltonian time-evolution,
   the STOQ protocol can also be used to compile gate sequences that approximately implement purely random unitaries in terms of an arbitrary gate set, without having any prior knowledge of the structure of the unitary.
   
\figureref{fig:random-unitary-compilation} shows
   typical results of repeatedly using the STOQ protocol
   to compile gate sequences for random two-qubit, three-qubit, and five-qubit unitaries,
   generated according to \cite{Mezzadri2006HowGroups},
   using a simple universal gate alphabet
     $G = \{ R_\varphi(\theta), XX(\theta) \}$.
   $R_\varphi(\theta)$ is a parameterized single-qubit rotation
      \begin{equation}
          \setlength\arraycolsep{2pt}
          R_\varphi(\theta) = 
            \begin{bmatrix}
                \cos \frac{\theta}{2}     & e^{-i(\frac{\pi}{2} + \varphi)} \sin  \frac{\theta}{2}  \\
                e^{-i(\frac{\pi}{2} - \varphi)} \sin  \frac{\theta}{2} & \cos \frac{\theta}{2}
            \end{bmatrix}
      \end{equation}
      with $0 \le \theta < 2\pi$ and $0 \le \varphi < 2\pi$.
      $XX(\theta)$ is a parameterized two-qubit entangling gate
      \begin{equation}
          \setlength\arraycolsep{2pt}
          XX(\theta) = 
            \begin{bmatrix}
                \cos \theta         & 0         & 0         & -i \sin \theta  \\
                0         & \cos \theta         & -i \sin \theta    & 0 \\
                0         & -i \sin \theta         & \cos \theta         & 0  \\
                -i \sin \theta         & 0         & 0         & \cos \theta 
            \end{bmatrix}
      \end{equation}
      with $0 \le \theta < 2\pi$.
      We note that the gate alphabet $G$ is a typical native gate set for trapped-ion quantum devices.

We observe that the final costs of compilation of these random unitaries are significantly larger than for compilation of the time-evolution unitaries discussed in \sectionref{sec:compilation-time-evolution}.
In particular,
    the final cost is approximately 0.1 for two-qubit random unitaries, 0.5 for three-qubit random unitaries, and 0.8 for five-qubit random unitaries.
    This indicates that the quality of the approximation for such random unitary compilations scales poorly with system size.
    This is not surprising, since reaching the vast majority of states in the Hilbert space of a system requires circuits of depth which grows exponentially with the dimension of the Hilbert space \cite{Knill1995ApproximationCircuits, Poulin2011QuantumSpace}.
    Nonetheless, the compilations generated by this method may be useful in scenarios where high-fidelity approximations are not required.
    
We also observe that the final cost of such random unitary compilations is relatively stable over a wide range of STOQ parameter values.
    Two primary parameters that can be adjusted in the STOQ algorithm in \figureref{fig:stoq-algorithm} are
        the annealing rate $\Delta\beta$,
            which is used to increment $\beta$ at each step inside the \texttt{IncreaseBeta} function,
        and the probability $p_\textrm{append}$ that the search appends a gate (as opposed to removing a gate) at each step,
            which occurs inside the \texttt{RandomChange} function.
    For compilation of three-qubit random unitaries,
        and for values $\Delta\beta \in \{0.001, 0.01, 0.1, 0.5\}$ and $p_\textrm{append} \in \{0.2, 0.5, 0.8\}$,
        we find that the average final cost remains between 0.398 (for $\Delta\beta=0.5$ and $p_\textrm{append}=0.2$) and 0.448 (for $\Delta\beta=0.001$ and $p_\textrm{append}=0.5$),
        where each pair of parameter values is averaged over 32 compilations using 100,000 iterations each.

To provide insight into the low-fidelity approximations of random unitaries produced by STOQ, we consider the case of target unitaries generated by random circuits of varying depth.
    To do this, we generate random five-qubit circuits of average depth ranging from 1 to 40, where the average depth is calculated as the total number of gates divided by the number of qubits.
    The rightmost plot in \figureref{fig:random-unitary-compilation} shows the final compilation cost after applying STOQ to generate an approximate compilation of the unitary corresponding to each random circuit.
    As might be expected, we observe that STOQ generates relatively high-fidelity approximations for shallow circuits, since such unitaries are known to be reachable with a fixed number of gates. But as the circuit depth increases, the resulting unitaries begin to look more like random unitaries, and the final compilation cost approaches that of the randomly-generated five-qubit unitary discussed previously.

\section{Discussion}

\subsection{Comparison with other methods}

We note that because the STOQ protocol requires calculating the product of the compiled sequence during each iteration, the computational cost of each iteration grows exponentially in the system size $n$. 

For compilation of time-evolution unitaries,
   this clearly means that STOQ will be less efficient in terms of runtime when compared to compilation methods based on product formulas,
   which in general have a computational cost
      that depends only on the number of terms in the Hamiltonian
      and is independent of the system size.

We note that unitaries generated via time evolution of a Hamiltonian often benefit from the sparsity of the Hamiltonian.
    In general, an $n$-qubit Hamiltonian has $4^n$ coefficients when expressed in the basis of Pauli operators.
    For the five-qubit version of the Hamiltonian in \equationref{eq:hamiltonian}, only nine of these 1024 coefficients are non-zero.
    Sparsity in the Hamiltonian greatly limits the subspace of the full operator space that can be reached by via time evolution, which in turn makes compilation a more feasible task and
    allows techniques such as Suzuki-Trotter and QDRIFT to be highly efficient.

Because the number of possible step proposals during each iteration of the STOQ search process is determined by the number of terms in the Hamiltonian, it is reasonable to infer that STOQ is similarly more effective when the problem structure contains such sparsity.
    This is further evidenced by the inability of the STOQ protocol to efficiently obtain low cost values when compiling sequences for random target unitaries, which are not sparse in general.

As demonstrated in this work, STOQ has some potential advantages for certain applications.
   One advantage is that repeated application of STOQ provides many independent approximate compilations of the same unitary.
      Each compilation creates a sequence that will cause the system state to traverse a different path in state space, some of which are remarkably different from the path that would be followed by deterministic product formula techniques.
      And as shown in this work, even stochastic techniques such as randomized Suzuki-Trotter or QDRIFT result in a compiled sequence that will cause the system state to follow very nearly the same path in state space as the deterministic version.
      
Another notable advantage of STOQ is that it generates meaningful results with arbitrary gate sets,
    since the protocol requires nothing of the gate set other than that the gates be unitary.

It is worth noting that STOQ is fundamentally different from existing gate-based randomized compilation techniques.
   In STOQ, the entire compilation is generated randomly, whereas in typical randomized compilation protocols, the process begins with an existing compilation of the desired unitary and adds local randomness in a manner that does not change the overall product of the sequence.

\subsection{Possible applications}

The capability of STOQ to independently generate many approximate compilations of a single unitary, particularly for unitaries corresponding to shallow random circuits, suggests that there may be practical applications of STOQ for tasks related to characterization of quantum devices.
Randomized benchmarking and related protocols also independently generate many compilations of the same unitary operation,
    but typically the unitary being compiled is just the identity operation, and the gate alphabet is usually the set of Clifford operations or some other non-universal gate set.
    These restrictions allow compilation to be efficient and exact.

Compilation with STOQ, on the other hand, can in principle be performed for any target unitary operation and with any gate alphabet.
The use of a stochastic compilation protocol similar to STOQ has been demonstrated to have potential advantages for characterization of analog quantum simulators \cite{Shaffer2021PracticalSimulators},
   in which many approximately-equivalent sequences are compiled and executed in order to assess the accuracy with which an analog quantum simulator has implemented the dynamics of the target Hamiltonian. We suggest that STOQ may also be useful for similar applications which do not require exact compilation, given that its requirements are less stringent than traditional protocols.

We also note that the cost function from \equationref{eq:cost}
    is similar to the one used in a proposed variational compilation algorithm called
        quantum-assisted quantum compiling \cite{Khatri2019Quantum-assistedCompiling}.
    In this scheme, in order to avoid the exponential runtime of evaluating the cost function classically, the cost function is evaluated on a quantum device.
Such an approach could in principle also be used to improve the scalability of STOQ.

\section{Conclusion}

This work has introduced STOQ, a stochastic search protocol for approximate unitary compilation into a sequence of gates from an arbitrary gate alphabet. We have described the procedure and details of its implementation, and we have demonstrated its performance by compiling time-evolution unitaries and random unitaries. We have also compared it to existing product-formula compilation techniques for time-evolution unitaries. We have observed that STOQ produces compilations that are less accurate than those produced by product-formula techniques, which indicates that STOQ is unlikely to be useful for applications that require high-fidelity compilation. We have noted that the primary advantage of STOQ is its ability to generate independent compilations that may cause the system to take significantly different paths through state space. This may be particularly useful for generating approximately-equivalent implementations of shallow random circuits for use on near-term quantum devices. Finally, we have discussed the applicability of STOQ to the area of device characterization, particularly for scenarios such as analog quantum simulation that cannot be covered by traditional randomized benchmarking techniques. We hope that STOQ may be a simple yet useful tool for exploring the performance and possible applications of near-term quantum devices.

\section*{Code Availability}

A Python implementation of STOQ is available at 
\url{https://github.com/rmshaffer/stoq-compiler}.

\section*{Acknowledgements}

The author thanks Hartmut H\"{a}ffner for many helpful discussions and inspiration for this work.

The author acknowledges government support under contract FA9550-11-C-0028
   and awarded by the Department of Defense,
   Air Force Office of Scientific Research,
   National Defense Science and Engineering Graduate (NDSEG) Fellowship, 32 CFR 168a.

%
%
\bibliographystyle{plainnat}
\bibliography{references}

\end{document}